\documentclass[12pt]{article}

\usepackage{amstex,amssymb,fullpage}
\numberwithin{equation}{section}

\begin{document}

\begin{titlepage}

\thispagestyle{empty}

\title{Asymptotic self-similarity breaking at\\
late times in cosmology.}

\author{J. Wainwright, $^\dag$\\
M. J. Hancock  $^\dag$\\
\&\\
C. Uggla $^\ddag$}

\date{October 17, 1998}

\maketitle
\vspace{5mm}

\begin{itemize}
\item[$^\dag$]
Department of Applied Mathematics\\
University of Waterloo\\
Waterloo,  Ontario \\
N2L 3G1\\
Canada

\item[$^\ddag$]
Department of Engineering Sciences, Physics and
Mathematics,\\
University of Karlstad,\\
S-65188 Karlstad,\\
Sweden.
\end{itemize}
\end{titlepage}

\newpage
\begin{abstract}
We study the late time evolution of a class of
exact anisotropic cosmological solutions of
Einstein's equations, namely spatially
homogeneous cosmologies of Bianchi type VII$_0$ 
with a perfect fluid source.  We show that, in
contrast to models of Bianchi type VII$_h$ 
which are asymptotically self-similar at late
times, Bianchi VII$_0$  models undergo a
complicated type of self-similarity breaking. 
This symmetry breaking affects the late time
isotropization that occurs in these models in a
significant way:  if the equation of state
parameter  $\gamma$  satisfies  $\gamma \leq
\frac{4}{3}$  the models isotropize as regards
the shear but not as regards the Weyl
curvature.  Indeed these models exhibit a new
dynamical feature that we refer to as {\it Weyl
curvature dominance}:  the Weyl curvature
dominates the dynamics at late times.  By
viewing the evolution from a dynamical systems
perspective we show that, despite the special
nature of the class of models under
consideration, this behaviour has implications
for more general models.
\end{abstract}

\section{Introduction}
In cosmology there are two asymptotic regimes,
namely

\begin{itemize}
\item[i)]
the approach to the initial singularity,
characterized by 
$l \rightarrow 0$, and

\item[ii)]
the late time evolution, characterized by  $l
\rightarrow \infty$,
\end{itemize}
\noindent
where  $l$  is the overall length scale.  In
other words, the asymptotic regimes correspond
to the extreme values of the length scale
variable. Alternatively, one can characterize
the asymptotic regimes in terms of the Hubble
variable 
$H$, which is related to  $l$  by
\begin{equation}
H = \frac{\dot{l}}{l}, \label{eq1.1}
\end{equation}
where the overdot denotes differentiation along
the fundamental congruence.   For
an ever-expanding universe,  $H > 0$  throughout
the evolution, and the singular regime is
characterized by  $H
\rightarrow +
\infty$  and the late-time regime by  $H
\rightarrow 0$.

The two asymptotic regimes are of course totally
different from a physical point of view, since
at the singularity physical quantities (for
example, the matter density, the shear of the
Hubble flow, the Weyl curvature) typically
diverge, while at late times physical quantities
typically tend to zero.  Despite this physical
distinction, the asymptotic regimes of the
simplest ever-expanding models, the isotropic
Friedmann-Lema\^{i}tre models  and the
anistropic but spatially homogeneous
models of Bianchi type I, have an important
property in common:  {\it they are asymptotically
self-similar} \footnote{The flat FL model is in
fact (exactly) self-similar (Eardley 1974, page
304).}, that is, they are approximated by
self-similar models in the asymptotic regime, as
was first pointed out by Eardley (1974, page
304).

The above behaviour becomes less surprising in
view of the following heuristic consideration: 
in both asymptotic regimes, one has a
self-gravitating system evolving in size through
many orders of magnitude (evolution into the past
for the singular regimes and into the future for
the late time regime), making it plausible that
the system might become scale-invariant
i.e. self-similar, asymptotically.  It is thus
tempting to speculate that a "Principle of
Asymptotic Self-Similarity" holds quite
generally for Bianchi universes, namely that any
Bianchi universe is approximated by a
self-similar Bianchi universe in the asymptotic
regimes.  However, the well-known Mixmaster
universes (Bianchi types VIII \& IX) provide a
decisive counter-example to the validity of such
a principle for the singular regime, since they 
oscillate indefinitely as the initial
singularity is approached into the past, and
thus do not have a well-defined asymptotic 
state.  Nevertheless the Mixmaster
universes reinforce the idea that self-similar
models are important as regards the overall
evolution of Bianchi universes, since they
evolve through an infinite sequence of Kasner
(i.e. self-similar) states as they approach the
initial singularity into the past.  Indeed,
research to date strongly suggests that a
non-tilted perfect fluid Bianchi universe either
approaches a unique self-similar asymptotic
state, or it passes through an infinite sequence
of self-similar states as one follows the
evolution into the past towards the initial
singularity.

The question remains as to the validity of the
Principle of Asymptotic Self-Similarity in the
late-time regime.  {\it Our main goal in this
paper is to show that the non-tilted Bianchi
VII$_0$  universes provide a counter-example to
asymptotic self-similarity in the late time
regime.}

One can gain a deeper understanding of
asymptotic self-similarity in Bianchi universes 
and of the reasons for its violation by viewing
the Einstein field equations from a dynamical
systems perspective.  In studying physical
phenomena, it is usually desirable to use
dimensionless variables.  In cosmology the
density parameter, a dimensionless quantity
defined by \footnote{From now on we will use
geometrized units, i. e.  $8 \pi G = 1$  and  $c
= 1$}
\begin{equation}
\Omega_0 = \biggl( \frac{8 \pi G \mu}{3H^2}
\label{eq1.5}
\biggr)_0 ,
\end{equation}
where  $\mu$  is the matter density,  $G$  is 
the gravitational constant and a subscript zero
means evaluation at the present time, is one of
the fundamental observational parameters.  In
addition the dimensionless quantity 
$(\sigma /H)_0$, where 
$\sigma$  is the shear scalar of the Hubble
flow, gives a measure of the anisotropy in the
Hubble flow.  This quantity is constrained by the
observations of the anisotropy in the
temperature of the cosmic microwave
background radiation (CMBR).

In view of these considerations, it is natural to formulate the Einstein field
equations using dimensionless variables that are
defined by normalizing in an appropriate way
with the Hubble scalar  $H$, as in Wainwright
and Hsu 1989.  The appropriate time variable is a
dimensionless quantity 
$\tau$  that is related to the length scale 
$l$  by 
\begin{equation}
l = l_0 e^{\tau}, \label{eq1.2}
\end{equation}
where  $l_0$  is a constant.   We shall refer
to  $\tau$  as {\it the dimensionless time
variable.}  In terms of 
$\tau$, the singular regime is defined by 
$\tau \rightarrow - \infty$  and the late time
regime for an ever-expanding model by $\tau
\rightarrow + \infty$.  The dimensionless time 
$\tau$  is related to clock time  $t$  by
\begin{equation}
\frac{dt}{d \tau} = \frac{1}{H}, \label{eq1.3}
\end{equation}
as follows from equations \eqref{eq1.1} and
\eqref{eq1.2}.  We refer to Wainwright and Ellis
1997 (Chapters 5 and 6) for full details \footnote{This work will henceforth be
referred to as WE.}.

For Bianchi universes the Einstein field
equations with a perfect fluid source and
an equation of state  $p = (\gamma -1) \mu$
 reduce to an autonomous system of ordinary
differential equations, thereby defining a
dynamical system.  It turns out that if one uses
the above dimensionless variables, the
equilibrium points (i.e. fixed points) of this
dynamical system correspond to self-similar
Bianchi universes, provided that  $\gamma > 0$. 
Within this framework, the evolution of a Bianchi
universe is described by an orbit of the
dynamical system, and asymptotic self-similarity
into the past/future means that the orbit
approaches an equilibrium point into the
past/future.  Conversely, {\it asymptotic
self-similarity is violated whenever the orbit
of a cosmological model does not approach a
single equilibrium point into the past/future}.

In the Mixmaster model the asymptotic
self-similarity breaking is due to the fact that
into the past, the orbits approach a bounded
two-dimensional attractor containing the set of
Kasner equilibrium points (the so-called
Mixmaster attractor, see WE, page 146), instead
of a single equilibrium point.  On the other
hand, we will show that the breaking of
asymptotic self-similarity in the late time
regime in Bianchi VII$_0$  models is due to the
fact that state space is unbounded, and the
orbits ``escape to infinity" at late times.  The
difference between the two types of symmetry
breaking is reflected in the behaviour of
dimensionless scalars formed by normalizing with
the Hubble scalar $H$.  In a self-similar model
such scalars are constant.   It follows that in
an asymptotically self-similar regime the limits
of such scalars exist.  In a Mixmaster universe
the limits of some scalars of this type do not
exist in the singular asymptotic regime (see WE,
figures 11.1-11.3), but all remain bounded.  On
the other hand, in a Bianchi VII$_0$  late time
regime we will show that, depending on the
equation of state of the perfect fluid, a
dimensionless scalar constructed from the Weyl
curvature tensor oscillates and becomes
unbounded.

In section 2 we introduce the three primary
dimensionless scalars and summarize their
behaviour in the late time asymptotic regime. 
This section contains no technical details, and
is intended to give an overview of the main
results.  In Section 3 we introduce the basic
dimensionless variables, which are based on the
orthonormal frame formalism, referring to
WE (chapters 5 and 6) for full details.  A change
of variable then leads to a new form of the
evolution equations that is adapted to the
oscillatory nature of the Bianchi VII$_0$ 
models.  The detailed asymptotic form of the
solutions of these evolution equations in the
late time regime is given (see theorems 3.1 -
3.3).  Section 4 contains a discussion of the
cosmological implications of our results.

There are three Appendices.  Appendix A contains
a proof of the fact that Bianchi VII$_0$ 
universes are not asymptotically self-similar at
late times, while Appendix B contains the
lengthy proof of Theorem 3.1, which gives the
asymptotic form of the solutions when the
equation of state parameter satisfies 
$\frac{2}{3} < \gamma < \frac{4}{3}$.  In
Appendix C we use the asymptotic expansions in
Theorems 3.1 - 3.3 to determine the asymptotic
form of the line-element, so as to be able to
compare our results with those of other
investigations.  We note, however, that
knowledge of the orthonormal frame components is
sufficient for analyzing any aspect of the
model, for example the degree of isotropization
and the anisotropy of the cosmic microwave
background radiation.  The asymptotic form of
the line-element is thus not central to our
conclusions, and hence is placed in an appendix.

\section{Main results}
Our main results concern the dynamics in the
late time regime of non-tilted perfect fluid
Bianchi universes of group type VII$_0$.  The
Bianchi VII$_0$ universes that are locally
rotationally  symmetric \footnote{See for
example WE, pages 22-3; we use the standard
abbreviation LRS.} either admit a group of
Bianchi type I or are flat FL universes, and
hence are asymptotically self-similar.  We thus
exclude this special class in the sequel.  The
first result concerns the breaking of asymptotic
self-similarity.
\vspace{3mm}

\noindent
{\bf Theorem 2.1:}\\
If  $\frac{2}{3} < \gamma < 2$, then any
non-tilted perfect fluid Bianchi universe of
group type  VII$_0$  that is not LRS is not
asymptotically self-similar as  $\tau \rightarrow
+ \infty$.\\

The proof of this Theorem, which requires
some technical results from the theory of
dynamical systems, is given in Appendix A.

As mentioned in the Introduction, the Hubble
scalar  $H$  is used to define dimensionless
variables in cosmology.  We recall that  $H$ is
defined by  $H = \frac{1}{3} u^a_{;a}$, where
{\bf u} is the 4-velocity of the Hubble flow
(the cosmological fluid), and determines the
lenth scale  $l$  according to \eqref{eq1.1}.  If
a Bianchi universe is asymptotically self-similar
at late times, then all dimensionless scalars
formed by normalizing with the Hubble scalar 
$H$ have limits as 
$\tau \rightarrow + \infty$.  As a result of
Theorem 2.1, it is possible that the limits of
some dimensionless scalars will not exist as 
$\tau \rightarrow + \infty$.  The next results
give the limiting behaviour of three
physically important dimensionless scalars. 

The first scalar is the {\it density
parameter}  $\Omega$, defined by \footnote{It is
customary to use a subscript zero on scalar such
as  $\Omega$  and  $H$, as in equation
\eqref{eq1.5}, to indicate their present day
value.}

\begin{equation}
\Omega = \frac{\mu}{3H^2}, \label{eq2.1}
\end{equation}
 where  $\mu$  is the matter density of the 
perfect fluid.  The density parameter gives a
measure of the dynamical significance of the
matter content of the model.  If  $\Omega \ll
1$  then the dynamics will be close to a vacuum
model, characterized by  $\Omega = 0$.

The second scalar is the {\it shear parameter} 
$\Sigma$  defined by
\begin{equation}
\Sigma^2 = \frac{\sigma^2}{3H^2}, \label{eq2.3}
\end{equation}
where \, $\sigma^2 = \tfrac{1}{2} \sigma_{ab}
\sigma^{ab},$ \,
and  $\sigma_{ab}$  is the rate-of-shear tensor
of the Hubble flow.  The shear parameter gives a
dimensionless measure of the anisotropy in the
Hubble flow by comparing the shear scalar 
$\sigma$  to the overall rate of expansion as
described by 
$H$.  The anisotropy in the temperature of the
CMBR enables one to estimate the value of 
$\Sigma$  at the present epoch (see for
example Maartens et. al. 1996).

The third scalar gives a dimensionless measure
of the Weyl curvature tensor, and is defined by 
\begin{equation}
\mathcal{W} = \frac{W}{H^2},
\label{eq2.4}
\end{equation}
where 
\begin{equation}
W^2 = \tfrac{1}{6}  \bigl( E_{ab}
E^{ab} + H_{ab} H^{ab} \bigr).
\label{eq2.5}
\end{equation}
Here  $E_{ab}$  and  $H_{ab}$  are the electric
and magnetic parts of the Weyl tensor, defined by
$$
E_{ab} = C_{arbs} u^r u^s, \quad H_{ab} = ^*
C_{arbs} u^r u^s,
$$
where  ${\bf u}$  is the 4-velocity of the
Hubble flow and $^*$  denotes the dual
operation.  We shall refer to  $\mathcal{W}$  as
{\it the Weyl parameter}.   $\mathcal{W}$  can
be regarded as describing the intrinsic
anisotropy in the gravitational field. 
Cosmological observations can in principle give
an upper bound on  $\mathcal{W}$, although
obtaining a strong bound is beyond the reach of
present day observations.  We refer specifically
to observations of the CMBR \footnote{The
analysis of Maartens et al 1995b does lead to a
strong upper bound on the Weyl parameter 
$\mathcal{W}$  (see their equation (41)). 
However, they make an assumption concerning the
time derivatives of the temperature harmonics
(assumption C2$'$) which is not observationally
testable at present.} (see Maartens et al
1995b), and observations of distant galaxies
(see Kristian \& Sachs 1966, page 398).  

The generalized Friedmann equation
\begin{equation}
3H^2 = \sigma^2 - \tfrac{1}{2} \, {^3R} + \mu,
\label{eq2.6}
\end{equation}
where  $^3R$  is the scalar curvature of the
spacelike hypersurfaces of homogeneity, places
bound on  $\Omega$  and  $\Sigma^2$.  Dividing
\eqref{eq2.6} by  $3H^2$ yields
\begin{equation}
1 = \Sigma^2 + K + \Omega, \label{eq2.7}
\end{equation}
where
$$
K = - \frac{^3R}{6H^2}
$$
is a dimensionless quantity.  Since  $K \geq 0$ 
for Bianchi type VII$_0$  geometries
\footnote{This result can be established using
equations (1.95), (1.103) and (1.105) in WE.},
equation \eqref{eq2.7} implies that 
\begin{equation}
\Sigma^2 \leq 1 \quad \text{and} \quad \Omega
\leq 1.
\label{eq2.8}
\end{equation}
On the other hand there is no general upper bound
for the Weyl parameter  $\mathcal{W}$.  Indeed 
we shall show that  {\em $\mathcal{W}$  is
unbounded in the state space of the Bianchi
 VII$_0$  cosmologies.} 

We now describe in turn the asymptotic behaviour
of the three primary expansion-normalized scalars
$\Omega, \Sigma$  and 
$\mathcal{W}$  in Bianchi VII$_0$  universes at
late times, i.e. as   $\tau \rightarrow +
\infty$.  These results will follow immediately
from Theorems 3.1 - 3.3 in Section 3.\\

\vspace{3mm}
\noindent
{\bf Theorem 2.2:}\\
For any non-tilted perfect fluid Bianchi
VII$_0$  universe that is not LRS, the density
parameter  $\Omega$  satisfies
\begin{equation*}
\lim_{\tau \rightarrow + \infty} \Omega = 
\begin{cases}
1 \qquad \quad,& \text{if} \quad \tfrac{2}{3} <
\gamma
\leq \tfrac{4}{3}\\[2mm]
\tfrac{3}{2} (2 - \gamma), \quad& \text{if} \quad
\tfrac{4}{3} < \gamma < 2.
\end{cases}
\end{equation*}

\vspace{3mm}
\noindent
{\bf Theorem 2.3:}\\
For any non-tilted perfect
fluid Bianchi VII$_0$  universe that is not
LRS, the shear parameter  $\Sigma$  satisfies
$$
\lim_{\tau \rightarrow + \infty} \Sigma^2 = 0,
\quad
\text{if} \quad \tfrac{2}{3} < \gamma \leq
\tfrac{4}{3}
$$
and
$$
\limsup_{\tau \rightarrow + \infty} \Sigma^2 =
\tfrac{1}{2} (3 \gamma -4), \quad  
\liminf\limits_{\tau
\rightarrow + \infty} \Sigma^2 = \tfrac{1}{4} (3
\gamma -4)^2, \quad \text{if} \quad \tfrac{4}{3}
< \gamma < 2.
$$

\noindent
{\bf Theorem 2.4:}\\
For any non-tilted perfect fluid Bianchi
VII$_0$  universe which is not LRS, the Weyl
curvature parameter  $\mathcal{W}$  satisfies
\begin{equation*}
\lim_{\tau \rightarrow + \infty} \mathcal{W}
\quad = 
\begin{cases}
0 \quad \quad ,& \text{if} \quad \tfrac{2}{3} <
\gamma < 1,\\[2mm]
L \neq 0,& \text{if} \quad \gamma = 1,\\[2mm]
+ \infty \quad ,& \text{if} \quad 1 <
\gamma < 2.
\end{cases}
\end{equation*}
where  $L$  is a constant that depends on the
initial conditions, and can have any positive
real value.

\noindent
The implications of Theorems 2.3 and 2.4 are as
follows:

\begin{itemize}
\item[i)]
there is a {\it shear bifurcation} at  $\gamma =
\tfrac{4}{3}$;  specifically, if $\gamma >
\tfrac{4}{3}$  the models no longer isotropize
as regards the shear of the Hubble flow, and in
fact  $ \, \lim\limits_{\tau \rightarrow +
\infty} \Sigma^2$  does not exist since
$\Sigma^2$ is oscillatory as  $\tau \rightarrow +
\infty$.

\item[ii)]
there is a {\it Weyl curvature bifurcation} at 
$\gamma = 1$;  specifically, if  $\gamma = 1$,
the models no longer isotropize as regards the
Weyl curvature, and if  $1 < \gamma < 2$  the
Weyl curvature moreover dominates the rate of
expansion at late times.
\end{itemize}
\vspace{3mm}
\noindent
{\it Comment:}  
By Theorems 2.2 - 2.4, the limits as 
$\tau \rightarrow +
\infty$  of  $\Omega, \Sigma$  and 
$\mathcal{W}$  exist if  $\tfrac{2}{3} < \gamma
\leq 1$.  One might thus expect that the models
are asymptotically self-similar in this case,
contradicting Theorem 2.1.  We will show in
Section 3, however, that if  $\tfrac{2}{3} <
\gamma \leq 1$, a dimensionless scalar
representing a time derivative of the Weyl
tensor is unbounded as \\
$\tau \rightarrow + \infty$.

\section{Evolution equations and asymptotic
behaviour}

In order to write the evolution equations for
the non-tilted perfect fluid Bianchi universes
of group type VII$_0$  in a suitable form, we use
the orthonormal-frame formalism, introduced in
cosmology by Ellis \& MacCallum 1969, in which
the commutation functions $\gamma^c_{ab}$ of the
orthonormal frame  $\{ {\bf e}_a \}$, defined by
\begin{equation*}
[{\bf e}_a,{\bf e}_b] =
\gamma^c_{ab}{\bf e}_c,
\end{equation*}
act as the basic variables of the gravitational
field.  We choose the frame to be invariant
under the isometry group G$_3$, and aligned with
the fluid velocity  ${\bf u}$   in the sense
that {\bf e}$_0=${\bf u}.  It follows that the
$\gamma^c_{ab}$ depend only on $t$, the clock
time along the fluid congruence.  For Bianchi
universes of group type VII$_0$  one can
specialize the frame so that the only non-zero
commutation functions are the Hubble scalar $H =
\tfrac{1}{3} \Theta$, where $\Theta$ is the rate
of expansion scalar, and the diagonal components
of the shear tensor
$\sigma_{\alpha \beta}$ and a matrix   $n_{\alpha
\beta}$,
\begin{equation}
\sigma_{\alpha \beta}= \text{diag} (\sigma_{11},
\sigma_{22}, \sigma_{33}), \quad n_{\alpha
\beta}= \text{diag}(0,n_2,n_3), \label{eq3.0}
\end{equation} 
with  $n_2 > 0$  and  $n_3 > 0$.
\noindent
(see Ellis \& MacCallum 1969). 
Since $\sigma_{\alpha
\beta}$ is trace-free, it has only two
independent components, which it is convenient
to label using
\begin{equation}
\sigma_+= \tfrac{1}{2} (\sigma_{22}+\sigma_{33}),
\quad \sigma_- = \tfrac {1}{2\sqrt{3}}
(\sigma_{22}- \sigma_{33}). 
\label{eq3.1}
\end{equation}
In analogy we label the components  $n_2, n_3$ by
\begin{equation}
n_+ = \tfrac{1}{2} (n_2 + n_3), \quad n_- =
\tfrac{1}{2 \sqrt{3}} (n_2 - n_3)
\label{eq3.2}
\end{equation}
The physical state of the models is thus
described by the variables
\begin{equation}
(H, \sigma_+, \sigma_-,n_+,n_-). \label{eq3.3}
\end{equation}

We now introduce expansion-normalized variables
and a dimensionless time  $\tau$  according to 
\begin{equation}
\Sigma_{\pm} = \frac{\sigma_{\pm}}{H}, \quad
N_{\pm} =  \frac{n_{\pm}}{H},
\label{eq3.4}
\end{equation}
and replace  $t$  by the dimensionless time
variable  $\tau$  according to equation (1.4)
(see WE, Chapter 5, for background and
motivation).  The evolution equation for  $H$ 
can now be written as 
\begin{equation}
H' = - (1+q) H,  \label{eq3.6}
\end{equation}
where  $'$  denotes differentiation with respect
to $\tau$, and  $q$  is the deceleration
parameter, given by
\begin{equation}
q = 2 \Sigma^2 + \tfrac{1}{2} (3 \gamma - 2)
\Omega. \label{eq3.7}
\end{equation}
(see WE, pages 113-4).  In addition  $\Sigma^2$ 
and  $\Omega$ are given by
\begin{equation}
\Sigma^2 = \Sigma_+^2 + \Sigma_-^2,
\label{eq3.8}
\end{equation}
\begin{equation}
\Omega = 1 - \Sigma_+^2 - \Sigma_-^2 - N_-^2.
\label{eq3.9}
\end{equation}

The evolution equations for  $\Sigma_{\pm}$ 
and  $N_{\pm}$  are
\begin{align}
\Sigma'_+ & = -(2-q) \Sigma_+ -2N^2_- \notag\\
\Sigma'_- & = -(2-q) \Sigma_- -2N_-N_+ \notag\\
N'_+ & = (q+2 \Sigma_+) N_+ + 6 \Sigma_- N_-
\label{eq3.10}\\
N'_- & = (q+2 \Sigma_+) N_- + 2 \Sigma_-
N_+. \notag
\end{align}
These equations follow from equations (6.9),
(6.10) and (6.35) in WE on setting  $N_1 =
0$.   For future reference we also note the
evolution equation for 
$\Omega$:
\begin{equation}
\Omega' = [2q - (3 \gamma -2)] \Omega, 
\label{eq3.11}
\end{equation}
(see WE, page 115).

The dimensionless state vector is  $(\Sigma_+,
\Sigma_-, N_+, N_-)$ and the physical region of 
state space is defined by the condition  $\Omega
\geq 0$, i.e.
\begin{equation}
\Sigma_+^2 + \Sigma_-^2 + N_-^2 \leq 1, 
\label{eq3.12}
\end{equation}
and the restrictions
\begin{equation}
N^2_+ - 3N^2_- > 0, \quad N_+ > 0, \label{eq3.13}
\end{equation}
which are a consequence of \eqref{eq3.2}, 
\eqref{eq3.4} and the restrictions  $n_2 > 0 ,
\, n_3 > 0$.   The variables  $\Sigma_+,
\Sigma_-$  and  $N_-$  are thus bounded, but
$N_+$  can take on any real value.  Indeed, we
show in Appendix A that if 
$\Omega > 0$  and  $\tfrac{2}{3} <
\gamma < 2$, then for any initial conditions,
\begin{equation}
\lim_{\tau \rightarrow + \infty} N_+ = +
\infty. \label{eq3.14} 
\end{equation}

The variable  $N_+$  does not appear in the
equation for  $\Sigma'_+$, but does play a
significant role in the equations for 
$\Sigma'_-$  and  $N'_-$, which are of the form

\begin{align*}
\Sigma'_-  & = - 2N_- N_+  + \{bounded \,
terms\}\\ 
N'_-  & = 2 \Sigma_- N_+  + \{bounded
 \, terms\}.
\end{align*}

The leading terms generate oscillations in 
$\Sigma_-$  and  $N_-$, and suggest that we
introduce polar coordinates in the  $\Sigma_-
N_-$-space.  We also replace the variable 
$N_+$  by 
$1/N_+$  so as to have a variable that
remains bounded as  $\tau \rightarrow +
\infty$.  We thus let
\begin{equation}
\Sigma_- = R \, \text{cos} \, \psi, \quad N_- =
R \, \text{sin} \, \psi,
\label{eq3.15}
\end{equation}
where  $R \geq 0$, and
\begin{equation}
M = \frac{1}{N_+}.
\label{eq3.16}
\end{equation}

A straightforward calculation using
\eqref{eq3.10}, \eqref{eq3.15} and \eqref{eq3.16}
yields the evolution equations for the new
variables  $(\Sigma_+, R, M, \psi)$  in the
following form:
\begin{align}
\Sigma'_+ & = - (2-Q) \Sigma_+ -R^2 +
(1+\Sigma_+) R^2 \, \text{cos} \, 2 \psi
\label{eq3.17}\\[3mm]
R' & = \bigl[ Q + \Sigma_+ -1 + (R^2 - 1 -
\Sigma_+) \, \text{cos} \, 2 \psi \bigr] R
\label{eq3.18}\\[3mm]
M' & = - \bigl[ Q + 2 \Sigma_+ + R^2 
( \text{cos} \, 2 \psi + 3 M \, \text{sin} \, 2
\psi) \bigr] M \label{eq3.19}\\[3mm]
\psi' & = \frac{1}{M} \bigl[2 + (1+ \Sigma_+) M
\, \text{sin} \, 2 \psi \bigr],  \label{eq3.20}
\end{align}
where
\begin{equation}
Q = 2 \Sigma^2_+ + R^2 + \tfrac{1}{2} (3 \gamma -
2) \Omega, \label{eq3.21}
\end{equation}
\begin{equation}
\Omega = 1 - \Sigma^2_+ - R^2 \label{eq3.22}
\end{equation}

In writing these equations we have expressed the
trigonometric dependence in terms of cos $2
\psi$  and sin $2 \psi$.  In particular, we have
decomposed the deceleration parameter  $q$, as
given by equation \eqref{eq3.7}, into a
non-oscillatory and an oscillatory part,
\begin{equation}
q = Q + R^2 \, \text{cos} \, 2 \psi.
\label{eq3.23}
\end{equation}

The variables  $\Sigma_+, R, M$  and  $\psi$ 
are required to satisfy the following
restrictions.  First, the requirement  $\Omega
\geq 0$  is equivalent to
\begin{equation}
\Sigma^2_+ + R^2 \leq 1.
\label{eq3.24}
\end{equation}
Second, the restrictions \eqref{eq3.13} in
conjunction with the definitions \eqref{eq3.15}
and \eqref{eq3.16} lead to
\begin{equation}
M > 0, \quad R \geq 0, \quad 3 R^2 M^2 \,
\text{sin}^2 \, \psi \leq 1 . \label{eq3.25}
\end{equation}
We note that if   $R = 0$  equation
\eqref{eq3.20} for  $\psi'$  becomes
irrelevant, and the remaining evolution
equations \eqref{eq3.17} and \eqref{eq3.19}
describe the LRS Bianchi VII$_0$  models (which
are equivalent to LRS Bianchi I models).

We now state the main results concerning the
asymptotic behaviour of the solutions of the DE
(3.17) - (3.20) with  $R> 0, \quad \Omega > 0
\quad \text{and} \quad \tfrac{2}{3} < \gamma <
2$.  For convenience, we introduce a new
parameter  $\beta$  given by
\begin{equation}
\beta = \tfrac{1}{2} (4 - 3 \gamma),
\label{eq3.26}
\end{equation}
and note the relations
\begin{equation}
\tfrac{3}{2} (2 - \gamma) = 1 + \beta, \quad
\tfrac{1}{2} (3 \gamma - 2) = 1 - \beta, \quad 3
(\gamma - 1) = 1 - 2 \beta. \label{eq3.27}
\end{equation}    
\vspace{3mm}

\noindent
{\bf Theorem 3.1:}\\
If  $\tfrac{2}{3} < \gamma < \tfrac{4}{3}$, any
solution of the DE \eqref{eq3.17} -
\eqref{eq3.20} with  $R > 0$  and  $\Omega > 0$ 
satisfies 
\begin{equation}
\lim_{\tau \rightarrow + \infty} M = 0 ,
\quad \lim_{\tau \rightarrow + \infty} R = 0
, \quad \lim_{\tau \rightarrow + \infty}
\Sigma_+ = 0  , \label{eq3.28}
\end{equation} 
and
\begin{align}
M  & = C_M e^{-(1-\beta) \tau} \Bigl[ 1 + O
\bigl( e^{-b \tau} \bigr) \Bigr]\notag\\
R  & = C_R e^{-\beta \tau} \Bigl[ 1 + O
\bigl( e^{-b \tau} \bigr) \Bigr] \label{eq3.29}\\
\Sigma_+  & = - \frac{C^2_R}{1-\beta} e^{-2 \beta
\tau} \Bigl[ 1 + O \bigl( e^{-b \tau} \bigr)
\Bigr], \notag
\end{align}
as  $\tau \rightarrow + \infty$, where  $\beta$ 
is given by \eqref{eq3.26}, 
$C_R$  and  $C_M$  are constants that depend on
the initial conditions, and  $b$  is a positive
constant.\\
\vspace{3mm}

\noindent
{\bf Theorem 3.2:}\\
If  $\gamma = \frac{4}{3}$, any solution of the
DE \eqref{eq3.17} - \eqref{eq3.20} with  $R >
0$  and  $\Omega
 > 0$  satisfies  \eqref{eq3.28}, and
\begin{align}
M  & = C_M \tau e^{- \tau} \Bigl[ 1 + O
\bigl( \tfrac{ln \tau}{\tau} \bigr)
\Bigr]\notag\\ R  & = \tfrac{1}{\sqrt{2}}
\tau^{- \frac{1}{2}}
\Bigl[ 1 + O \bigl(\frac{ln \tau}{\tau} \bigr)
\Bigr] \label{eq3.30}\\
\Sigma_+  & = - \tfrac{1}{2} \tau^{-1}
\Bigl[ 1 + O \bigl(\frac{ln \tau}{\tau} \bigr)
\Bigr], \notag
\end{align}
where  $C_M$  is a constant that depends on the
initial conditions.\\
\vspace{3mm}

\noindent
{\bf Theorem 3.3:}\\
If  $\tfrac{4}{3} < \gamma < 2$, any solution of
the DE \eqref{eq3.17} - \eqref{eq3.20} with  $R
> 0$  and  $\Omega > 0$  satisfies

\begin{equation}
\lim_{\tau \rightarrow + \infty} M = 0 \quad ,
\quad \lim_{\tau \rightarrow + \infty} \Sigma_+
 = \beta \quad , \quad \lim_{\tau \rightarrow +
\infty} R^2 = - \beta (1 + \beta) \quad ,
\label{eq3.31}
\end{equation} 
with
\begin{equation}
M = C_M e^{-(1+\beta) \tau} \Bigl[ 1 + O \bigl(
e^{-b \tau} \bigr) \Bigr], \label{eq3.32}
\end{equation}
where  $\beta$  is given by \eqref{eq3.26}, 
$C_M$  is a constant that depends on the initial
conditions and  $b$  is a positive constant.\\

Theorems 3.1 - 3.3 enable us to determine the
asymptotic behaviour of any physical or
geometrical quantity in a non-tilted Bianchi
VII$_0$  universe, and in particular, provide
the proofs for Theorems 2.2 - 2.4.  The shear
parameter  $\Sigma$  is obtained from equations
\eqref{eq3.8} and \eqref{eq3.15}, and the density
parameter  $\Omega$  from equation
\eqref{eq3.22}:
\begin{equation}
\Sigma^2 = \Sigma^2_+ + R^2 \, \text{cos}^2
\psi, \quad \Omega = 1 - \Sigma^2_+ - R^2.
\label{eq3.33}
\end{equation}
Theorems 3.1-3.3, in conjunction with equation
\eqref{eq3.33}, now lead directly to Theorems
2.2 and 2.3.

Next, the asymptotic form of the Hubble
parameter  $H$  is determined algebraically,
since the density parameter  $\Omega =
\mu/(3H^2)$  has a non-zero limit as  $\tau
\rightarrow + \infty$, for  $\tfrac{2}{3} <
\gamma < 2$, and the matter density is given by 
$\mu = \mu_0 (l/l_0)^{- 3 \gamma}$, or
equivalently,  $\mu = \mu_0 e^{- 3 \gamma
\tau}$, as follows from the contracted Bianchi
identities.  Knowing  $H$, the relation between
clock time  $t$  and dimensionless time  $\tau$ 
can be obtained by integrating equation
\eqref{eq1.3}.  On introducing the parameter 
$\beta$, we obtain

\begin{equation}
H \approx H_0 e^{-(2-\beta)\tau}, \quad H_0 t
\approx \frac{1}{2-\beta} e^{(2-\beta)\tau},
\label{eq3.34}
\end{equation}
as  $\tau \rightarrow + \infty$.  The asymptotic
expansions in Theorems 3.1-3.3 also determine
the asymptotic form of the line-element by a
quadrature and by algebraic means  (See Appendix
C).

We now discuss the behaviour of the Weyl
curvature.  
We require an explicit expression for the Weyl
parameter  $\mathcal{W}$, defined by equations
\eqref{eq2.4} and \eqref{eq2.5}.  For non-tilted
Bianchi VII$_0$  models,  $E_{ab}$  and 
$H_{ab}$, and their dimensionless counterparts
\begin{equation}
\mathcal{E}_{ab} = \frac{E_{ab}}{H^2} \quad ,
\quad \mathcal{H}_{ab} = \frac{H_{ab}}{H^2} ,
\label{eq3.35}
\end{equation}
are diagonal relative to the standard
orthonormal frame (see WE, Chapter 6,
Appendix).  In analogy with \eqref{eq3.1} we
define
\begin{align}
\mathcal{E}_+  = \tfrac{1}{2} (\mathcal{E}_{22
} + \mathcal{E}_{33}) \quad , \quad
\mathcal{E}_- & = \tfrac{1}{2 \sqrt{3}}
(\mathcal{E}_{22} - \mathcal{E}_{33})
\notag \\
\mathcal{H}_+  = \tfrac{1}{2} (\mathcal{H}_{22
} + \mathcal{H}_{33}) \quad , \quad
\mathcal{H}_- & = \tfrac{1}{2 \sqrt{3}}
(\mathcal{H}_{22} - \mathcal{H}_{33}).
\label{eq3.36} 
\end{align}
It then follows from equations \eqref{eq2.4} and
\eqref{eq2.5} that
\begin{equation}
\mathcal{W}^2 = \mathcal{E}^2_+ +
\mathcal{E}^2_- + \mathcal{H}^2_+ +
\mathcal{H}^2_- . 
\label{eq3.37}
\end{equation}

In terms of the variables  $\Sigma_+, R, M,
\psi$  as defined by equations \eqref{eq3.15} and
\eqref{eq3.16}, the components 
$\mathcal{E}_{\pm}$  and 
$\mathcal{H}_{\pm}$  are given by
\begin{align}
\mathcal{E}_+ & = \Sigma_+ (1 + \Sigma_+) +
\tfrac{1}{2} R^2 (1-3  \text{cos}  2 \psi)
\notag \\
\mathcal{H}_+ & = - \tfrac{3}{2} R^2  \text{sin} 
2 \psi  \notag\\
\mathcal{E}_- & = \tfrac{2R}{M} \bigl[
\text{sin} 
\psi + \tfrac{1}{2} M (1-2 \Sigma_+)  \text{cos} 
 \psi \bigr]  \label{eq3.38}\\
\mathcal{H}_- & = \tfrac{2R}{M} \bigl[-
\text{cos} \, \psi  -
\tfrac{3}{2} M \Sigma_+  \text{sin}  \psi \bigr].
\notag 
\end{align}
These expressions follow from equations (6.36)
and (6.37) in WE.  Since  $\Sigma_+$  and  $R$ 
are bounded, it follows from (3.37) and
\eqref{eq3.38} that 
\begin{equation}
\mathcal{W} = \frac{2R}{M} [1 + O(M)], \quad
\text{as} \quad \tau \rightarrow +
\infty. \label{eq3.39}
\end{equation}

This result, in conjunction with
Theorems 3.1-3.3, yields
\begin{equation}
\mathcal{W}^2 =
\begin{cases}
4 \bigl( \frac{C_R}{C_M} \bigr)^2
e^{2(1-2\beta)\tau} \bigl[ 1 + O (e^{-b\tau})
\bigr], &  \text{if} \quad \tfrac{2}{3} < \gamma
< \tfrac{4}{3} \\[2mm]
\frac{2}{C^2_M} \tau^{-3} e^{2 \tau} \bigl[ 1 +
O (\frac{ln \tau}{\tau}) \bigr], & \text{if}
\quad \gamma = \tfrac{4}{3} \\[2mm]
\frac{-4 \beta (1+\beta)}{C^2_M}
e^{2(1+2\beta)\tau} \bigl[ 1 + O (e^{-b\tau})
\bigr], &  \text{if} \quad \tfrac{4}{3} < \gamma
< 2. \label{eq3.40}
\end{cases}
\end{equation} 
Theorem 2.4 follows immediately from this
result using equation
\eqref{eq3.27}.  We note that  $\mathcal{W}$  
diverges most rapidly for  $\gamma
= \tfrac{4}{3}$.

As mentioned at the end of Section 2, if 
$\tfrac{2}{3} < \gamma \leq 1$  the limits of
the primary dimensionless scalars  $\Sigma^2,
\Omega$  and  $\mathcal{W}$  exist, even though
the model is not asymptotically self-similar. 
We can use the asymptotic expressions in Theorem
3.1 to show that {\it if}  $\tfrac{2}{3} < \gamma
\leq 1$, {\it there is an
expansion-normalized scalar that is
unbounded as  $\tau \rightarrow +
\infty$.}  We need to know the dominant term in the
derivatives of  $\mathcal{E}^{(n)}_-$  and 
$\mathcal{H}_-^{(n)}$, where  $(n)$  denotes
the  $n^{th}$  derivative with respect to 
$\tau$.  One can calculate 
$\mathcal{E}_-^{(1)}$ and  $\mathcal{H}_-^{(1)}$
by differentiating equations \eqref{eq3.38} with
respect to  $\tau$, using the evolution
equations \eqref{eq3.17} - \eqref{eq3.20}.  The
dominant term arises from differentiating the
leading trigonometric function, and we obtain
\begin{align}
\mathcal{E}_-^{(1)} & = \tfrac{4R}{M^2} [
\text{cos} \, \psi + O (M)], \notag \\
\mathcal{H}_-^{(1)} & = \tfrac{4R}{M^2} [
\text{sin} \, \psi + O (M)], 
\label{eq3.41}
\end{align}
as  $\tau \rightarrow + \infty$.  Repeating this
process shows that the dominant term in 
$\mathcal{E}_-^{(n)}$  and 
$\mathcal{H}_-^{(n)}$  grows like  $R/M^{n+1}$ 
as  $\tau \rightarrow + \infty$.
  Theorem 3.1 implies that
$$
\frac{R}{M^{n+1}} \, \sim \,
e^{[n(1-\beta)-(2\beta-1)]\tau} \quad \text{as}
\quad \tau \rightarrow + \infty.
$$
It follows from this result that 
$\mathcal{E}_-^{(n)}$  and 
$\mathcal{H}_-^{(n)}$  are unbounded as  $\tau
\rightarrow + \infty$  if and only if \\ $n > 6(1
- \gamma)/(3\gamma-2)$.  Thus, for any 
$\gamma$  in the range  $\tfrac{2}{3} < \gamma
\leq 1, \, \, \mathcal{E}^{(n)}_-$  and 
$\mathcal{H}_-^{(n)}$  are unbounded as  $\tau
\rightarrow + \infty$  for  $n$  sufficiently
large.

We conclude this Section by giving a heuristic
justification of Theorems 3.1 - 3.3.  The key
result underlying the proof is that 
\begin{equation}
\lim_{\tau \rightarrow + \infty} M = 0,
\label{eq3.42}
\end{equation}
as follows from equations \eqref{eq3.14} and
\eqref{eq3.16}.  We shall show that in fact  $M$ 
tends to zero at an exponential rate.  Equation
\eqref{eq3.20} implies that once  $M$  is close
to zero,  $\psi'$  is large, and that  $\psi$ 
grows exponentially as  $\tau \rightarrow +
\infty$.  Because of this exponential growth,
{\it the solutions of the evolution equations
\eqref{eq3.17} - \eqref{eq3.20} are approximated
by solutions of the simplified DE obtained by
dropping the cos  $2 \psi$  and sin  $2 \psi$ 
terms in equations \eqref{eq3.17} and
\eqref{eq3.18}}:
\begin{align}
\hat{\Sigma}'_+ & = -(2 - \hat{Q})
\hat{\Sigma}_+ - \hat{R}^2 \label{eq3.43}\\
\hat{R}' & =  (\hat{Q} + \hat{\Sigma}_+ -1)
\hat{R}, \notag
\end{align}
where we use the symbol  $\hat{~}$ to distinguish
the variables in \eqref{eq3.43} from those in
equations \eqref{eq3.17}, \eqref{eq3.18} and
\eqref{eq3.21}.  This behaviour, which is shown
very convincingly in numerical experiments, is
due to the fact that the cos  $2 \psi$  and sin 
$2 \psi$  terms are essentially nullified since
they change sign repeatedly over an increasingly
short time scale.

The state space for the variables 
$\hat{\Sigma}_+$  and  $\hat{R}$  is 
$$
\hat{\Sigma}_+^2 + \hat{R}^2 \leq 1, \quad
\hat{R} \geq 0
$$
(see \eqref{eq3.24} and \eqref{eq3.25}).  Since
this set is closed and bounded we can use
standard methods to analyze the asymptotic
behaviour of solutions of the DE
\eqref{eq3.43}.  We find that if  $\tfrac{2}{3}
< \gamma \leq \tfrac{4}{3}$  the point 
$(\hat{\Sigma}_+, \hat{R}) = (0,0)$  is a global
sink for solutions with  $\Omega > 0$, and if 
$\tfrac{4}{3} < \gamma < 2$, the point 
$(\hat{\Sigma}_+, \hat{R}) = (\beta, \sqrt{-
\beta (1+\beta)} )$  is a global sink for
solutions with  $\Omega > 0$  and  $R > 0$. 
One can also determine the asymptotic form of
the solutions of the simplified DE
\eqref{eq3.43}, which gives precisely the
leading asymptotic dependence on  $\tau$  in
Theorem 3.1 - 3.3.

In summary, for any initial conditions, equation
\eqref{eq3.42} implies that  $M$  will be close
to zero for  $\tau$  sufficiently large, after
which the evolution will be governed
asymptotically by the simplified DE
\eqref{eq3.43}.  Proving the theorems thus
hinges on proving that the oscillatory terms do
in fact become negligible for  $\tau$ 
sufficiently large.  A detailed proof of
Theorem 3.1 is given in Appendix B.  The proof
of Theorem 3.3 ( the case  $\tfrac{4}{3} <
\gamma < 2$) is of a similar nature to that of
Theorem 3.1, while the proof of Theorem 3.2
(the case of  $\gamma = \tfrac{4}{3}$) is more
complicated, since the global sink for the
averaged DE (3.43) is non-hyperbolic.  The
proof thus requires the use of cente manifold
theory (see for example Carr 1981).  The
details of this case will be published
elsewhere.

\section{Discussion}
In this paper we have described for the first
time the evolution into the future of any
non-tilted Bianchi VII$_0$  universe with
perfect fluid matter content and equation of
state \\ $p = (\gamma - 1) \mu$, for  $\gamma$ 
in the range  $\frac{2}{3} < \gamma < 2$.  We
have also given asymptotic forms as  $\tau
\rightarrow + \infty$  of various physical and
geometrical quantities of interest.  Although
certain aspects of this problem have been
studied before, most notably by \footnote{We
discuss the results of these authors in
Appendix C, when we describe the asymptotic
form of the line-element.} Collins \& Hawking
1973a, Doroshkevich et al 1973 and Barrow \&
Sonoda 1986, a general treatment has not been
given, and moreover a number of important
properties of the dynamics have remained
unnoticed. 

The most significant feature of the late time
evolution is the breaking of asymptotic
self-similarity (Theorem 2.1), characterized by
oscillations that become increasingly rapid in
terms of dimensionless time  $\tau$  as $\tau
\rightarrow + \infty$.  This oscillatory
behaviour leads to the phenomenon of {\it Weyl
curvature dominance}, i.e.
expansion-normalized scalars constructed from
the Weyl tensor become unbounded as  $\tau
\rightarrow + \infty$.  We note that this type
of asymptotic symmetry breaking is quite
different from that displayed by the Mixmaster
models in the singular asymptotic regime, which
has a stochastic nature but with all
expansion-normalized scalars remaining bounded.

It is worth noting that both types of asymptotic
self-similarity breaking create difficulties
when one attempts to solve the Einstein field
equations numerically.  In the Bianchi VII$_0$
case, if one uses 
$\tau$  as the time variable the increasingly
rapid oscillations (like cos ($e^{\lambda
\tau}$)) cause numerical difficulties, while if
one changes to a time variable relative to which
the oscillations are asymptotically periodic
(e.g. conformal time 
$\eta$; see table C.1) then the overall
evolution slows to such an extent that one
cannot integrate very far into the late time
asymptotic regime.  On the other hand, as one
follows the evolution of a Mixmaster model into
the past, the model spends an increasingly long
time ($\tau$-time) in the successive Kasner
states so that it is likewise difficult to
integrate far into the singular asymptotic
regime.  \footnote{We refer to Berger et al 1997
for a new algorithm for improving the situation
somewhat.}

The late time behaviour of the Bianchi VII$_0$ 
models also has interesting implications as
regards the question of isotropization.  In a
spatially homogeneous cosmology, there are two
physical manifestations of anisotropy,
\begin{itemize}
\item[i)]
the {\it shear} of the timelike congruence that
represents the large scale distribution of the
matter,
\item[ii)]
the {\it Weyl curvature}, which can be viewed as
describing the intrinsic anisotropy in the
gravitational field:  it determines up to
four preferred directions, the principal null
directions of the gravitational field.
\end{itemize}
\noindent
The shear parameter  $\Sigma$  and the
Weyl parameter $\mathcal{W}$  quantify these
anisotropies.

In discussions of isotropization in the
literature restrictions are imposed on 
$\Sigma$, namely
\begin{equation}
\lim_{\tau \rightarrow + \infty} \Sigma = 0,
\label{eq4.1}
\end{equation}
corresponding to {\it asymptotic
isotropization}  (see for example, Collins and
Hawking 1973b page 324), or
\begin{equation}
\Sigma \ll 1 ,
\label{eq4.2}
\end{equation}
over some finite time interval \footnote{For
example, the time elapsed since last scattering
or the time elapsed since nucleosynthesis.},
corresponding to {\it intermediate
isotropization} (see Zeldovich and Novikov 1983,
page 550, and more recently, Wainwright et al
1998, pages 331-2 and 343-4).  Restrictions are
not usually imposed on the Weyl curvature.  For
the class of non-tilted Bianchi models  $\Sigma
= 0$  characterizes the FL models and hence
implies that  $\mathcal{W} = 0$, making it
tempting to conjecture that
\eqref{eq4.1} and
\eqref{eq4.2} imply the corresponding results
for the Weyl parameter  $\mathcal{W}$.  Our
analysis shows decisively that this conjecture
is false:  even though \eqref{eq4.1} holds, the
Weyl parameter 
$\mathcal{W}$ can be arbitrarily large for dust 
$(\gamma = 1)$  and is unbounded for the
parameter range  $1 < \gamma \leq \frac{4}{3}$ 
(which includes radiation), as  $\tau \rightarrow
+ \infty$.  The implication is that {\it a model
which satisfies
\eqref{eq4.1} or \eqref{eq4.2} is not
necessarily close to isotropy, since 
$\mathcal{W}$  may not be small.}

It is of interest to compare the late time
evolution of Bianchi VII$_0$  universes with the
evolution of perturbations of the flat FL
model.  We can do this by linearizing the
Bianchi VII$_0$  evolution equations about the
FL model, given by  $\Sigma_+ = 0 = R$.  Since
the perturbed solutions are spatially
homogeneous and have zero vorticity, it follows
that the scalar and vector perturbations
are zero, giving purely tensor
perturbations (Goode 1989, Section V).  The exact
solutions of the linearized equations involve
Bessel functions but for our purposes it is
sufficient to know the asymptotic forms,
\begin{equation}
\Sigma_+^{lin} \approx \quad C_{\Sigma}
e^{-(1+\beta) \tau} \quad , \quad R_{lin}
\approx \quad C_R e^{-\beta \tau} \quad , \quad
M_{lin} \approx \quad C_M e^{-(1-\beta)\tau}.
\label{eq4.3}
\end{equation}
Comparing this result with Theorem 3.1, we see
that {\it the linearized equations predict the
correct asymptotic form of  $R$  and  $M$, for 
$\gamma$  in the range}  $\frac{2}{3} < \gamma <
\frac{4}{3}$.  This implies that solutions of
the linearized equations lead to the same
asymptotic form for the Weyl parameter 
$\mathcal{W}$  as the exact equations (see
equation \eqref{eq3.39}).  Thus, {\it the Weyl
curvature bifurcation at 
$\gamma = 1$  is a property of the linearized
solutions as well as the exact solutions.}

On the other hand, it follows from equation
\eqref{eq4.3} and Theorem 3.1 that the
linearized equations predict that  $\Sigma_+$ 
decays faster than it actually does, i.e.
the asymptotic form of  $\Sigma_+$  is a
non-linear effect.  Furthermore, comparing
equations \eqref{eq4.3} with Theorem 3.2, we see
that the asymptotic form of all variables in the
case of radiation  $(\gamma = \frac{4}{3})$ is a
non-linear phenomenon.

We now discuss the implications of our results
as regards the temperature of the CMBR.  We have
seen that the shear of the Hubble flow, which is
described by the two frame components 
$\Sigma_{\pm}$, has two noteworthy features, in
particular for the case of dust  $(\gamma = 1):$

\begin{itemize}
\item[i)]
the  $\Sigma_-$  component oscillates
increasingly rapidly while tending to zero, as
the dimensionless time 
$\tau \rightarrow + \infty$,

\item[ii)]
the rate of decay of the  $\Sigma_+$ component
is a non-linear effect.
\end{itemize}
The temperature of the CMBR is given by the
formula \footnote{In papers analysing the
anisotropy of the temperature of the CMBR the
linearized version of this formula, valid for  
$\Sigma \alpha \beta << 1$, is usually given. 
See for example Collins \& Hawking 1973a.}
\begin{equation}
T_o = T_e exp \biggl[ -
\int^{\tau_{o}}_{\tau_{e}} (1 + \Sigma_{\alpha
\beta} K^{\alpha} K^{\beta}) d \tau \biggr],
\label{eq4.4}
\end{equation}
where the  $K^{\alpha}$  are the
direction cosines of a particular null geodesic, 
$\Sigma_{\alpha
\beta} = \sigma_{\alpha \beta} / H$  is the
dimensionless shear tensor, and  $\tau$  is the
dimensionless time.  The subscripts  $o$  and
$e$  refer to the present time and the time of
emission, respectively.  It is thus evident that
the link between the temperature anistotropy and
the shear is complicated, particularly in view
of the oscillatory nature of the shear, and that
determining a reliable bound on the shear
scalar  $\Sigma$  will necessitate integrating
the full evolution equations and null geodesic
equations numerically \footnote{The anisotropy
in the temperature of the CMBR in Bianchi
VII$_0$ models has been investigated by
Collins and Hawking 1973a and by Doroshkevich
et al 1975.  In contrast to Collins and
Hawking, Doroshkevich et al argue that it is not
sufficient to use linear perturbations of the FL
models.}.  Without investigating this link
further, we can, however, use the fact that
small 
$\Sigma$  does not imply small  $\mathcal{W}$  to
make the following assertion:  {\it a highly
isotropic CMBR temperature at the present time
does not lead to restrictions on the Weyl
parameter.}  Indeed {\it the Bianchi VII$_0$
universes provide an example of cosmological
models in which the CMBR temperature is highly
isotropic but the Weyl parameter is not small.} 
As mentioned in Section 3, the analysis of
Maartens et al 1995b shows that in order to
obtain a bound on  $\mathcal{W}$  one requires an
observational bound on the {\it time
derivatives} of the temperature harmonics, which
is not attainable in practice.

We conclude with some remarks interpreting our
results from a broader perspective.  Despite the
special nature of the models under consideration
- they are spatially homogoneous and
moreover are not of the most general Bianchi
type - the complicated dynamical behaviour that
we have described may have broad significance,
for the following reason.  The four-dimensional
state space of the non-tilted Bianchi VII$_0$ 
models is an invariant subset of the state space
for more general models, and by continuity the
orbits in an open subset of the full state space
will ``shadow", i.e. approximate, the orbits in
the invariant subset.  In other words, {\it an
open subset of more general classes of models
will display the features of Bianchi VII$_0$ 
models}, at least during a finite time
interval.  Thus, for example, we can assert
that there exist Bianchi VII$_0$  models with
tilt, and Bianchi VIII or IX models with or
without tilt, with the property that during some
extended epoch the shear parameter  $\Sigma$  is
small but the Weyl curvature parameter 
$\mathcal{W}$ is large.  More generally, this
property will also be displayed by spatially
inhomogeneous cosmologies, for example  $G_2$ 
cosmologies
\footnote{As an example of the dynamics of a
special class of models occurring in a more
general class (but for the singular asymptotic
regime), we note that Weaver et al 1998 have
recently shown numerically that Mixmaster-like
oscillations do in fact occur in 
$G_2$-cosmologies in the presence of a
magnetic field, generalizing the behaviour of
the magnetic Bianchi VI$_0$  cosmologies, whose
state space is an invariant subset of the
infinite dimensional state space of the
$G_2$-cosmologies (see Leblanc et al 1995)},
which contain Bianchi VII$_0$  cosmologies as a
special case.  Based on our experience with the
hierarchy of Bianchi models, namely that
dynamical complexity increases with the
dimension of the state space, we expect the
dynamics of Bianchi models to provide a lower
bound, so to speak, for the dynamical complexity
of spatially inhomogeneous models. 
Understanding the dynamics of Bianchi universes
is thus a necessary first step in the study of
more general models.

\section*{Appendix A:  Proof of Theorem 2.1}

We consider the state space  $S$  of the Bianchi
VII$_0$  models defined by the inequalities 
\begin{equation}
N_+ >0, \quad N_+^2 - 3N_-^2 > 0, \quad \Omega >
0,
\tag{A.1} \label{eqA.1}
\end{equation}
(see \eqref{eq3.12} and \eqref{eq3.13}).  The
function

\begin{equation}
Z = \frac{(N_+^2 -3N_-^2)^v  \, \Omega}{(1 + v
\Sigma_+)^{2(1+v)}}, \quad \text{with}
\quad v = \tfrac{1}{4} (3 \gamma - 2),\tag{A.2}
\label{eqA.2}
\end{equation}
satisfies  $0 < Z < + \infty$  on  $S$, and the
evolution equations \eqref{eq3.10} -
\eqref{eq3.11} imply that
\begin{equation}
\frac{Z'}{Z} =  \quad \frac{4 [(\Sigma_+ + v)^2
+ (1 - v^2) \Sigma_-^2]}{1 + v \Sigma_+}.
\tag{A.3} \label{eqA.3}
\end{equation}
Thus if  $\tfrac{2}{3} \leq \gamma \leq 2, \quad
Z$  is increasing along orbits in  $S$, since  $0
\leq v \leq 1$.

The set of boundary points of  $S$  that are not
contained in  $S$  is the set  $\overline{S}/S$,
where 
$\overline{S}$  is the closure of  $S$.  It
follows from \eqref{eqA.1} that 
$\overline{S}/S$  is defined by one or both of
the following equalities holding:
\begin{equation}
\Omega = 0 \quad , \quad N^2_+ - 3N_-^2 = 0.
\notag
\end{equation}

By equation \eqref{eqA.2}, $Z$  is defined and
equal to zero on  $\overline{S}/S$.  We can now
apply the Monotonicity Principle (see WE,
Theorem 4.12, with ``decreasing" replaced by
``increasing") to conclude that for any
point $ x \in S$, the
 $\omega$-limit set $ \omega ( x)$
is contained in the subset of $\overline{S}/S$ 
that satisfies the condition 
${\lim\limits_{ y \rightarrow s}} Z (y) \neq 0$,
where  
$s \in \overline{S} / S$ 
and 
$ y \in S$.  This subset is
the empty set, since  $Z = 0$  on 
$\overline{S}/S$, we conclude that  $\omega (x)
= \phi$  for all  $x \in S$.  

We now prove that 
\begin{equation}
\lim_{\tau \rightarrow + \infty} N_+ = + \infty
\tag{A.4} \label{eqA.4}
\end{equation}
for each orbit in  $S$.  Suppose that 
\eqref{eqA.4} does not hold.  Then there exists
a number  $b > 0$  such that for any  $\tau_0$,
there exists a 
$\tau > \tau_0$  with  $N_+ (\tau) < b$.  Since
the other variables are bounded, this implies
that the orbit  $x (\tau)$  has
infinitely many points in a compact subset of 
$S \subset \mathbb{R}^4$  and hence has a limit
point in  $S$, contradicting  $\omega x) =
\phi$.  Thus \eqref{eqA.4} holds.  In
conclusion, we have proved that no orbit in 
$S$  is future asymptotic to an equilibrium
point in  $S$  and that for every orbit in  $S$,
\eqref{eqA.4} holds.

\section*{Appendix B:  Proof of Theorem 3.1}
The proof of Theorems 3.1 - 3.3 involves two
main steps:
\begin{itemize}
\item[i)]
Prove that
\begin{equation*}
\lim_{\tau \rightarrow + \infty} (\Sigma_+, R) = 
\begin{cases}
(0,0), \quad & \tfrac{2}{3} < \gamma \leq
\tfrac{4}{3} \\
(\beta, \sqrt{- \beta (1 + \beta)}), \quad &
\tfrac{4}{3} < \gamma < 2
\end{cases}
\end{equation*}
for all initial states.

\item[ii)]
Deduce the asymptotic form of  $(\Sigma_+, R,
M)$  as  $\tau \rightarrow + \infty$.
\end{itemize}

The evolution equations \eqref{eq3.17}
- \eqref{eq3.20} are of the form
\begin{align}
x' & = f_0 (x) + f_1 (x) \, \text{cos} \, 2 \psi
\notag \\
M' & =  \bigl[ m_0 (x) + m_1 (x) \, \text{cos}
\, 2 \psi + m_2 (x) M \, \text{sin} \, 2
\psi \bigr] M \tag{B.1} \label{eqB.1}\\
\psi' & = \tfrac{2}{M} \bigl[1 + s(x) M \,
\text{sin} \, \psi \bigr],
\notag 
\end{align}
where the arbitrary functions are polynomials
in  $x = (\Sigma_+, R)$.  We recall equation
\eqref{eq3.42}, namely,  ${\lim\limits_{\tau
\rightarrow + \infty}} M = 0$.

In the analysis we will encounter scalar
functions 
$$
Z = Z (x,M),
$$  
whose evolution equation is of the form 
\begin{equation}
Z' = \bigl[ z_0 (x) + z_1 (x) \, \text{cos} \, 2
\psi + z_2 (x) M \, \text{sin} \, 2 \psi \bigr]
Z, \tag{B.2} \label{eqB.2}
\end{equation}
where  $z_0, z_1$  and  $z_2$  are polynomials,
and
\begin{equation}
z_0 (0) = - \eta < 0,
\tag{B.3} \label{eqB.3}
\end{equation}
It is understood that  $x, M$  and  $\psi$  in
\eqref{eqB.2} satisfy the DE \eqref{eqB.1}.

Suppose it is known that all solutions of the
DE \eqref{eqB.1} satisfy
$$
H_1: \quad \lim_{\tau \rightarrow + \infty} x
(\tau) = 0,
$$
or the stronger condition
$$
H_2: \quad x (\tau) = O (e^{-b \tau}), \quad M =
O(e^{-b \tau}) \quad \text{as} \quad \tau
\rightarrow + \infty,
$$
$$
\text{for some constant} \quad b > 0.
$$
These conditions in fact imply that  $Z$  tends
to zero at an exponential rate.  The desired
conclusions, given in terms of the constant 
$\eta$  in \eqref{eqB.3}, are as follows,
\begin{align*}
C_1 & : \quad Z = O \bigl(e^{(- \eta + \delta)
\tau}
\bigr)  \quad \text{as} \quad \tau \rightarrow +
\infty, \text{for any} \quad \delta > 0,\\
C_2 & : \quad Z = C_z e^{- \eta \tau} \bigl[ 1 +
O (e^{- b \tau}) \bigr], \quad \text{as}
\quad
\tau
\rightarrow + \infty,\\
& \text{where  $C_z$  is a constant that depends
on the initial condition.}
\end{align*}
\noindent

\vspace{3mm}

\noindent
{\bf Proposition B.1:}\\
$$
H_1 \quad  \text{implies} \quad  C_1 \quad
\text{and} \quad H_2 \quad  \text{implies} \quad 
C_2.
$$

\noindent
{\it Proof:}  The idea is to rescale  $Z$ 
so as to "suppress" the rapidly oscillating
term  $z_1 (x) \, \text{cos} \, 2 \psi$   in
(B.2) which may not tend to zero as  $\tau
\rightarrow +
\infty$.\\
Let
\begin{equation}
\overline{Z} = \frac{Z}{1 + \frac{1}{4} M z_1
(x) \, \text{sin} \, 2 \psi}
\tag{B.4} \label{eqB.4}
\end{equation}
Differentiating \eqref{eqB.4} with respect to
 $\tau$, and using \eqref{eqB.1} and
\eqref{eqB.2} leads to an equation of the form
\begin{equation}
\overline{Z}' = [z_0 (x) + MB] \overline{Z},
\tag{B.5} \label{eqB.5}
\end{equation}
where  $B = B (x, M, \, \text{cos} \, 2 \psi, \,
\text{sin} \, 2 \psi)$  is bounded for
sufficiently large  $\tau$.  

Suppose that  $H_1$ holds.  Since  $M$  tends to 
$0$  as  $\tau \rightarrow + \infty$, it follows
from
\eqref{eqB.3} that given  $\delta > 0$,
$$
z_0 (x) + M B \leq - \eta + \delta,
$$
for all  $\tau$  sufficiently large, leading to
$C_1$.

Suppose that  $H_2$  holds.  Then \eqref{eqB.5}
can be written in the form
$$
\overline{Z}' = \bigl[ - \eta + O (e^{-b \tau})
\bigr] \overline{Z}
$$
as  $\tau \rightarrow + \infty$, and  $C_2$ 
follows. \hspace{4.15in}  $\Box$ \\

We shall also encounter evolution equations of
the form 
\begin{equation}
Z' = - \eta Z + B,
\tag{B.6} \label{eqB.6}
\end{equation}
where  $\eta$  is a constant and  $B = B (x,M,
\text{cos} \, 2 \psi, \, \text{sin} \, 2 \psi)
$  is bounded.  By multiplying by  $e^{- \eta
(\tau - \tau_0)}$  and integrating from 
$\tau_0$  to $\tau$,  this DE can be written
in the equivalent integral form
\begin{equation}
Z (\tau) = e^{- \eta (\tau - \tau_0)} Z (\tau_0)
+ \int^{\tau}_{\tau_0} e^{- \eta (\tau - s)} B
ds, \tag{B.7} \label{eqB.7}
\end{equation}
\noindent
where the arguments of  $B$  are evaluated at 
$s$.  In connection with this integral form, we
will need \\

\noindent
{\bf Gronwall's Lemma:}\\
If  $v(\tau)$  is a non-negative  $C^1$ 
function, and 
$$
v (\tau) \leq C + \delta \int^{\tau}_{\tau_0} v
(s) ds,
$$
where  $C, \delta$  are positive constants, then
$$
v (t) \leq C e^{\delta (\tau - \tau_0)} \quad
\text{for all} \quad \tau \geq \tau_0.
$$
\noindent
{\it Proof:}  See for example, Hirsch \&
Smale (1974), page 169.  \\[3mm]
We can now give the proof of Theorem 3.1.\\

\noindent
{\bf {\em Proof of Theorem 3.1:}}\\[2mm]
We first prove that all solutions of the DE
\eqref{eqC.1} satisfy  ${\lim\limits_{\tau
\rightarrow +
\infty}} x (\tau) = 0$.  We write the evolution
equation \eqref{eq3.11} for  $\Omega$  in the
form
$$
\Omega' = 2 \bigl[ (1 + \beta) \Sigma^2_+ +
\beta R^2 + R^2 \, \text{cos} \, 2 \psi \bigr]
\Omega,
$$
using \eqref{eq3.21} - \eqref{eq3.23} and
\eqref{eq3.27}.

In analogy with \eqref{eqB.4}, we let
\begin{equation}
\overline{\Omega} = \frac{\Omega}{1 +
\frac{1}{2} M R^2 \, \text{sin} \, 2 \psi}.
\tag{B.8} \label{eqB.8}
\end{equation}
Differentiating with respect to  $\tau$  and
using \eqref{eq3.18} - \eqref{eq3.20} leads to
an equation of the form
\begin{equation}
\overline{\Omega}' = 2 \bigl[ (1 + \beta)
\Sigma^2_+ + R^2 (\beta + M B) \bigr]
\overline{\Omega},
\tag{B.9} \label{eqB.9}
\end{equation}
where  $B$  is a bounded expression in 
$\Sigma_+, R, M, \, \text{cos} \, 2 \psi$  and
sin  $2 \psi$, for  $\tau$  sufficiently large. 
Since  ${\lim\limits_{ \tau \rightarrow +
\infty}} M = 0$  and  $\beta > 0$, it follows
from
\eqref{eqB.9} that there exists  $\tau_0$  such
that  $\overline{\Omega}' \geq 0$  for all 
$\tau > \tau_0$.  Since  $\Omega \leq 1$,
equation \eqref{eqB.8} and the bound
\eqref{eq3.25} imply that  $\overline{\Omega}$ 
is bounded above.  It thus follows that 
${\lim\limits_{ \tau \rightarrow +
\infty}} \overline{\Omega}$  exists, say
\begin{equation}
\lim_{\tau \rightarrow + \infty}
\overline{\Omega} = L,
\tag{B.10} \label{eqB.10}
\end{equation}
which, by equations \eqref{eqB.8}, and
\eqref{eq3.42} gives  ${\lim\limits_{ \tau
\rightarrow +
\infty}} \Omega = L$.  Since  $\Omega \leq 1$,
we have  $L \leq 1$.  Suppose  $L < 1$.  Then
for  $\tau$  sufficiently large, it follows
that  $\Sigma^2_+ + R^2 = 1 - \Omega >
\frac{1}{2} (1 - L)$,\\
Equation \eqref{eqB.9} now implies
$$
\frac{\overline{\Omega}'}{\overline{\Omega}} >
\beta (1 - L) + 2 \Sigma^2_+ + R^2 M B.
$$
Since ${\lim\limits_{ \tau \rightarrow +
\infty}} M = 0$  we have  $|R^2 M B | <
\frac{\beta}{2} (1 - L)$  for  $\tau$ 
sufficiently large.  It follows that  
$$
\frac{\overline{\Omega}'}{\overline{\Omega}} >
\tfrac{1}{2} \beta (1 - L) > 0
$$ 
for  $\tau$  sufficiently large, contradicting 
\eqref{eqB.10}.  Thus  $L = 1$, i.e. 
${\lim\limits_{\tau \rightarrow + \infty}}
\Omega = 1$, which implies
$$
\lim_{\tau \rightarrow + \infty} (\Sigma^2_+ +
R^2) = 0, \quad \text{i.e.} \quad \lim_{\tau
\rightarrow + \infty} x (\tau) = 0.
$$

We can now use Proposition B.1, with 
$H_1$  satisfied, to show that  $R$  and  $M$ 
decay exponentially to zero.  The evolution
equations \eqref{eq3.18} and \eqref{eq3.19} are
of the form \eqref{eqB.2} with the constant 
$\eta$  in \eqref{eqB.3} given by 
\begin{equation*}
\eta  = - 2 \beta \quad \text{for} \quad R^2,
\quad
\eta  = - (1 - \beta) \quad \text{for} \quad M.
\end{equation*}
The proposition thus implies that 
\begin{equation}
R^2 = O \bigl( e^{(-2 \beta + \delta) \tau}
\bigr), \quad M = O \bigl( e^{(-(1- \beta) +
\delta) \tau} \bigr), 
\tag{B.11} \label{eqB.11}
\end{equation}
as  $\tau \rightarrow + \infty$.

The evolution equation \eqref{eq3.17} for 
$\Sigma_+$  can be written in the form
\begin{equation}
\Sigma'_+ = - (1 + \beta) \Omega \Sigma_+ - R^2
(1 + \Sigma_+) (1 - \, \text{cos} \, 2 \psi)
\tag{B.12} \label{eqB.12}
\end{equation}
using \eqref{eq3.21}, \eqref{eq3.22} and
\eqref{eq3.26}.  Further rearrangement yields
$$
\Sigma'_+ = - (1 + \beta) \Sigma_+ + B (x, \,
\text{cos} \, 2 \psi),
$$
where
\begin{equation}
|B| \quad \leq \quad (1 + \beta)
|\Sigma_+|^3 + c_1 R^2, \tag{B.13}
\label{eqB.13}
\end{equation}
with  $c_1$  a constant.  By \eqref{eqB.7}, this
DE has the equivalent integral form
\begin{equation}
\Sigma_+ (\tau) = e^{-(1 + \beta) (\tau -
\tau_0)} \Sigma_+ (\tau_0) +
\int^{\tau}_{\tau_0} e^{-(1 + \beta) (\tau - s)}
B d s
\tag{B.14} \label{eqB.14}
\end{equation}
where the arguments of  $B$  are evaluated at 
$s$.  By equation \eqref{eqB.11},

$$
R^2 < c_2 e^{(- 2 \beta + \delta) \tau} \quad
\text{for} \quad \tau > \tau_0.
$$
Since  ${\lim\limits_{\tau \rightarrow + \infty}}
\Sigma_+ = 0$  by  $H_1$, we can choose 
$\tau_0$  large enough that 
$$
\Sigma^2_+ < \frac{\delta}{1 + \beta} \quad
\text{for} \quad \tau > \tau_0.
$$
Using these inequalities with \eqref{eqB.13}, it
follows from equation \eqref{eqB.14} that
\footnote{The inequality  $-(1+ \beta) < - 2 \beta +
\delta$  is used to make all constants in the
exponentials the same.} 

$$
|\Sigma_+( \tau)| \quad  \leq \quad Ce^{(-2
\beta+ \delta) \tau} \quad + \quad \delta
\int^{\tau}_{\tau_0} e^{ (-2 \beta+
\delta)(\tau-s)}|  \Sigma_+ (s)| ds,
$$
Multiplying by  $e^{(2 \beta - \delta) \tau}$ 
yields
$$
v (\tau) \quad \leq \quad C + \delta
\int^{\tau}_{\tau_0} v (s) ds,
$$
where  $v(\tau) = e^{(2 \beta - \delta) \tau}
|\Sigma_+ (\tau)|$.  Gronwall's lemma thus
implies that 
$$
|\Sigma_+ (\tau)| \quad \leq \quad C e^{(-2
\beta + 2 \delta) \tau} \quad \text{for} \quad
\tau \geq \tau_0.
$$

At this stage we have shown that hypothesis 
$H_2$  is satisfied.  We can thus use
Proposition \eqref{eqB.1} to conclude that 
\begin{equation}
R = C_R e^{- \beta \tau} \bigl[ 1 + 0 (e^{-
\beta \tau}) \bigr] , \quad M = C_M
e^{-(1-\beta) \tau} \bigl[ 1 + 0 (e^{- b \tau}
) \bigr], \tag{B.15} \label{eqB.15}
\end{equation}
as  $\tau \rightarrow + \infty$  (see the
discussion leading to equation \eqref{eqB.11}.

The final step is to deduce the asymptotic form
of  $\Sigma_+ (\tau)$.  It is necessary to do a
change of variable in order to suppress the
rapidly oscillating term cos  $2 \psi$  in
\eqref{eqB.12}.  Define 

\begin{equation}
\overline{\Sigma}_+ = \Sigma_+ - \tfrac{1}{4} M
R^2 (1 + \Sigma_+) \, \text{sin} \, 2 \psi.
\tag{B.16} \label{eqB.16}
\end{equation}
Differentiating \eqref{eqB.16} with respect to 
$\tau$, and using  \eqref{eqB.1} and
\eqref{eqB.2} leads to an equation of the form
\begin{equation}
\Sigma'_+ = - (1 + \beta) \overline{\Sigma}_+ -
R^2 + \overline{B}, \tag{B.17} \label{eqB.17}
\end{equation}
where
$$
\overline{B} = (1 + \beta) \Sigma^3_+ + \beta
\Sigma_+ R^2 + M R^2 B,
$$
and  $B$  is a complicated expression that is
bounded as  $\tau \rightarrow + \infty$.  By
\eqref{eqB.7}, the DE \eqref{eqB.17} has the
equivalent integral form
\begin{equation}
\overline{\Sigma}_+ (\tau) = e^{-(1+ \beta)(\tau
- \tau_0)} \Sigma_+ (\tau_0)
- \int^{\tau}_{\tau_0} e^{-(1+ \beta)(\tau-s)}
R^2 ds - \int^{\tau}_{\tau_0} e^{-(1+
\beta)(\tau-s)}
\overline{B} ds.\tag{B.18} \label{eqB.18}
\end{equation}

The rates of decay for  $\Sigma_+, R$  and  $M$ 
obtained so far imply that the integral
involving  $R^2$  is the dominant term as  $\tau
\rightarrow + \infty$.  Using \eqref{eqB.15},
this integral can be evaluated, giving
\begin{equation}
\int^{\tau}_{\tau_0} e^{-(1+ \beta)(\tau-s)} R^2
ds = \frac{C^2_R}{1 - \beta}
e^{-2 \beta \tau} \bigl[ 1 + O (e^{- \overline{b}
\tau})
\bigr],
\tag{B.19} \label{eqB.19}
\end{equation}
as  $\tau \rightarrow + \infty$, for some
positive constant  $\overline{b}$.  The
asymptotic form for  $\Sigma_+ (\tau)$,
$$
\Sigma_+ (\tau) = - \frac{C^2_R}{1 - \beta}
e^{-2 \beta \tau} \bigl[ 1 + O (e^{-b \tau})
\bigr],
$$
as  $\tau \rightarrow + \infty$, follows from
\eqref{eqB.16}, \eqref{eqB.18} and
\eqref{eqB.19}, on noting that the term  $M
R^2$  in \eqref{eqB.17} is of order 
$e^{-(1+\beta) \tau}$  as  $\tau \rightarrow +
\infty$.  The value of the positive constant 
$b$  in the error term is re-defined as
necessary.

\section*{Appendix C:  The metric approach}
Other investigations of Bianchi VII$_0$ 
universes have used metric tensor components as
basic variables.  We can obtain the asymptotic
behaviour of the metric components directly from
our results in Section 3, and we now do this for
the purpose of comparison.

For any Bianchi model one can introduce a set of
group-invariant and time-independent one-forms 
$W^{\alpha}, \alpha = 1, 2, 3$, 
relative to which the line-element has the form
\begin{equation}
ds^2 = -dt^2 + g_{ \alpha \beta} (t)
W^{\alpha} W^{\beta},
\tag{C.1} \label{eqC.1}
\end{equation}
where  $t$  is clock time along the normal
congruence of the group orbits (e.g. WE, page
39).  For group type VII$_0$, the one-forms can
be chosen to satisfy
\begin{equation}
d W^1 = 0,  \quad  d W^2 =
W^3 \wedge W^1,  \quad d W^3 = W^1 \wedge W^2.
\tag{C.2} \label{eqC.2}
\end{equation}

If the matter-energy content is a non-tilted
perfect fluid, then  $g_{\alpha \beta} (t)$  can
be diagonalized, and for our purposes it is
convenient to  use the Misner labelling of the
metric components.
\begin{equation}
ds^2 = -dt^2 + l^2 \bigl[ e^{-4 \beta^+} 
(W^1)^2 + e^{2 (\beta^+ + \sqrt{3}
\beta^-)} (W^2)^2 + e^{2(\beta^+ -
\sqrt{3} \beta^-)} (W^3)^2 \bigr]
\tag{C.3} \label{eqC.3}
\end{equation}
(Misner 1969, page 1323), where  $l$  is the
length scale function.  The structure relations
are invariant under a rescaling \footnote{This
transformation is an element of the
automorphism group for Bianchi VII$_0$ (see, for
example, Jantzen 1984).} 
$W^2 \rightarrow \lambda W^2, \quad W^3
\rightarrow
\lambda W^3$, where  $\lambda$  is a constant,
which implies that there is freedom in
redefining  $l$   and  $\beta^+$  according to 
\begin{equation}
l \rightarrow e^{2C} l, \quad \beta^+
\rightarrow \beta^+ + C,
\tag{C.4} \label{eqC.4}
\end{equation}
where  $C$  is a constant.

The relations between the metric variables and
the orthonormal frame variables are given in
WE (Chapter 9), for Bianchi models of class
A.  Specializing equations (10.5) and (10.44) in
WE to Bianchi VII$_0$  and using (10.7), (10.11),
 (6.8) and (6.35) in WE leads to 
\begin{equation}
N_+ + \sqrt{3} N_- = \frac{1}{Hl} e^{2(\beta^+ +
\sqrt{3}
\beta^-)}, \quad N_+ - \sqrt{3} N_- =
\frac{1}{Hl} e^{2(\beta^+ - \sqrt{3} \beta^-)}.
\tag{C.5} \label{eqC.5}
\end{equation}
Introducing the variables $R, M$ and  $\psi$  by
equations \eqref{eq3.15} and \eqref{eq3.16} gives
\begin{equation}
\tanh (2 \sqrt{3} \beta^-) =  \sqrt{3} R M \,
\text{sin} \, \psi,
\tag{C.6} \label{eqC.6}
\end{equation}

\begin{equation}
e^{4 \beta^+} = \frac{l^2 H^2}{M^2} ( 1
- 3 R^2 M^2 \, \text{sin}^2  \psi).
\tag{C.7} \label{eqC.7}
\end{equation}
In addition, WE (10.46) reads
\begin{equation}
\Sigma_{\pm} = \frac{d \beta^{\pm}}{d \tau},
\tag{C.8} \label{eqC.8}
\end{equation}
where  $\tau$  is the dimensionless time
variable defined by equation \eqref{eq1.3}.  We
note that the length scale  $l$  is related to 
$\tau$  by equation \eqref{eq1.2}

We can now use the asymptotic expressions for 
$\Sigma_+, R, M$  and  $\psi$  in Theorems
3.1 - 3.3 to find the asymptotic form of 
$\beta^+$  and  $\beta^-$, and the relation
between clock time  $t$  and dimensionless time 
$\tau$.  Since  $M
\rightarrow 0$  as  $\tau \rightarrow + \infty$,
equation \eqref{eqC.6} implies that  $\beta^-
\rightarrow 0$  as  $\tau \rightarrow 0$,  and
that for  $|\beta^-| \ll 1$  we have  $\beta^-
\approx \frac{1}{2} R M \, \text{sin} \, \psi$,
which determines  $\beta^-$  directly.  We
integrate \eqref{eqC.8} to find  $\beta^+$  up to
an additive constant.  

By using the freedom \eqref{eqC.4} in
conjunction with equations \eqref{eq1.2},
\eqref{eq3.34} and \eqref{eqC.7} we can arrange
that the constant $C_M$  in the asymptotic form
of  $M$  in Section 3 is related to  $H_0$  and 
$l_0$  by
\begin{equation}
C_M = H_0 l_0.
\tag{C.9} \label{eqC.9}
\end{equation}

We now give the asymptotic expressions for 
$\beta^+$  and  $\beta^-$  as  $\tau \rightarrow
+ \infty$  in the three cases.

\begin{itemize}
\item[i)] \quad $\tfrac{2}{3} < \gamma <
\tfrac{4}{3} :$
\begin{equation}
\beta^+  \approx \frac{C_R^2}{2 \beta (1 -
\beta)} e^{-2 \beta \tau}, \quad \beta^- \approx
\tfrac{1}{2} C_R H_0 l_0 e^{- \tau} \, \text{sin}
\, [ \psi (\tau)], \tag{C.10} \label{eqC.10}
\end{equation}
with 
$$
 \psi (\tau) \approx \frac{2}{H_0
l_0 (1 - \beta)} e^{(1 - \beta) \tau} + \psi_0
$$

\item[ii)] \quad $\gamma = \tfrac{4}{3} :$ 
\begin{equation}
\beta^+  \approx - \tfrac{1}{2} l n \tau, \quad
\beta^- \approx \tfrac{1}{2 \sqrt{2}} H_0 l_0
\tau^{\tfrac{1}{2}} e^{- \tau} \, \text{sin}
\, [ \psi (\tau)], \tag{C.11} \label{eqC.11}
\end{equation}
with 
$$
 \psi' (\tau) \approx \frac{2}{H_0
l_0} \, \frac{e^{\tau}}{\tau}.
$$
In this case the asymptotic form of  $\psi
(\tau)$  cannot be expressed in terms of
elementary functions.

\item[iii)] \quad $ \tfrac{4}{3} < \gamma < 2 :$
\begin{equation}
\beta^+  \approx - \beta \tau, \quad
\beta^- \approx  \sqrt{- \beta (1+ \beta)} 
e^{- (1 + \beta) \tau} \, \text{sin}
\, [ \psi (\tau)], \tag{C.12} \label{eqC.12}
\end{equation}
with 
$$
 \psi (\tau) \approx \frac{2}{H_0
l_0 (1+ \beta)} \, e^{(1+ \beta) \tau} + \psi_0.
$$
\end{itemize}

We now briefly describe the results concerning
the late-time asymptotic behaviour of Bianchi
VII$_0$  universes that have been given in the
literature.  Comparing the results is made
difficult by the fact that a variety of
different time variables have been used.  For
convenience we summarize the time variables in
Table C.1.  (see also WE, page 244).  The
asymptotic dependence follows from
\vspace{4mm}

\begin{center}
\begin{tabular}{|c|c|c|c|}
\hline
Name & Definition & Values & Dependence on
$\tau$  as  $\tau \rightarrow + \infty$ \\ \hline
$~$ & $~$ & $~$ & $~$\\[1mm]
Clock time  $t$  &  $\frac{dt}{d \tau} =
\frac{1}{H}$  &  $ 0 < t < + \infty$  
&  $H_0 t \approx \frac{1}{2-\beta} e^{(2-\beta)
\tau.}$
\\ [3mm] BKL \, $T$  &  $\frac{dt}{dT} = l^3$ 
&  
$-
\infty < T < 0$   &   $H_0 l_0^3 T \approx-
\frac{1}{1+ \beta}e^{-(1+ \beta) \tau}$  \\[3mm]
Conformal $\eta$  &  $\frac{dt}{d \eta} = l$  & 
$0 < \eta < + \infty$  &  $H_0 l_0 \eta \approx
\frac{1}{1- \beta} e^{(1- \beta) \tau}$ \\ [3mm]
$G_2$-adapted  $\xi$  &   $\frac{dt}{d \xi} = l
e^{-2 \beta^+}$  &   $0 < \xi < + \infty$  &  
$\xi \approx \eta$ \, if  $\frac{2}{3} < \gamma <
\frac{4}{3}.$ \\ [3mm] \hline
\end{tabular} 
\end{center}

\begin{quote}
{\footnotesize Table C.1  Time variables for
Bianchi VII$_0$  universes.  Dimensionless time 
$\tau$  assumes all real values,  $- \infty <
\tau < + \infty$,  and  $\tau \rightarrow +
\infty$  in the late time asymptotic regime.  The
constant $\beta$  is given by  $\beta
= \frac{1}{2} (4 - 3 \gamma)$, where  $\gamma$ 
is the equation of state parameter.}
\end{quote}

\noindent
equations \eqref{eq3.34} and \eqref{eq1.2} and
the fact that  $\beta^+
\rightarrow 0$  as  $\tau \rightarrow + \infty$ 
if  $\frac{2}{3} < \gamma < \frac{4}{3}$  (see
\eqref{eqC.10}).  The  $\xi$-time was apparently
first used for Bianchi cosmologies by Collins \&
Hawking (1973a, see page 328), and subsequently
by Siklos (1980).  We use the name 
``$G_2$-adapted time" because the 
$G_3$  isometry group admits a two-parameter
Abelian subgroup  $G_2$, and $l e^{-2 \beta^+}$ 
is the length scale in the preferred direction
orthogonal to the orbits of the  $G_2$.

Collins and Hawking 1973a analyzed the
asymptotic evolution of Bianchi VII$_0$ 
universes assuming that the matter in the model
consisted of two non-interacting components,
dust $(\gamma = 1)$  and radiation  $(\gamma =
\frac{4}{3})$.  The dust dominates at late
times.  The tilt was not assumed to be zero. 
They proved that if the initial state is close
enough to the Einstein-de Sitter model, then the
Bianchi VII$_0$  model isotropizes in the sense
that  $\sigma/H \rightarrow 0$  and 
$\beta^{\pm} \rightarrow \beta^{\pm}_\infty$ 
as  $t \rightarrow + \infty$, where 
$\beta^{\pm}_\infty$  are constants (see their
Theorem 3).  Prior to proving this result they
heuristically determine the asymptotic form of 
$\beta^{\pm}$, finding that

$$
\beta^- \approx e^{- \alpha} (A \text{cos}  2 \xi
+ B \text{sin}  2 \xi), \quad \beta^+ \approx C
e^{-
\alpha},
$$
(see page 332).  To facilitate comparison we
note that  $e^{\alpha} = l, \quad H =
\dot{\alpha}$, and in their line-element in
Section IV, $\lambda = - 2 \beta^+, \quad
\lambda = 2 \sqrt{3} \beta^-, \quad \psi = b_1 =
b_2 = 0$.  On using Table C.1 and noting that $l
= l_0 e^{\tau}$, we see that their result agrees
with our result \eqref{eqC.10}, when  $ \gamma =
1$  i.e.  $\beta = \frac{1}{2}$.  It is of
interest that if the field equations are
linearized about the Einstein  de-Sitter model
they do not give the correct asymptotic form
for  $\beta^+$.  This fact can be seen most
clearly from the subsequent paper Collins and
Hawking (1973b) in which they give the
asymptotic form of the solution of the linear
field equations for Bianchi VII$_0$  universes
with dust.  In terms of the conformal time 
$\eta$  (see Table C.1),

$$
\beta_{11} = - 2 \beta^+ \approx
\frac{A}{\eta^3},
\quad \beta_{22} - \beta_{33} = 2 \sqrt{3}
\beta^- \approx \frac{1}{\eta^2} (B \text{cos} 
2 \eta + C \text{sin}  2 \eta)
$$
(see pages 316-7).

Doroshkevich et al 1973 analyzed the non-tilted
Bianchi VII$_0$  universes using the
line-element \eqref{eqC.3} and the BKL time 
$T$.  For dust  $(\gamma = 1)$, they give the
asymptotic form

$$
\beta^-  \approx C_1 T^{-\frac{2}{3}} \,
\text{sin} \, (C_2 T^{-\frac{1}{3}} + \psi_0),
\quad l^2 e^{-2 \beta_+}  \approx C_3
T^{-\frac{4}{3}}, 
$$
(see Appendix I, equation (I.20)).  To
facilitate comparison we note that their metric
variables  $\gamma, \mu$  and  $(\lambda_1
\lambda_2)^{\frac{1}{2}}$   are related to ours
by  $\gamma = l^6, \mu = 4 \sqrt{3} \beta^-$ 
and  $(\lambda_1 \lambda_2)^{\frac{1}{2}} = l^2
e^{-2 \beta^+}$. On using Table C.1 we see that
their result for  $\beta^-$ agrees with our
result \eqref{eqC.10} when  $\gamma = 1$  (i.e.
$\beta = \frac{1}{2}$).  Their result for 
$\beta^+$  simply confirms that  $e^{\beta^+}
\rightarrow$  constant as  $t \rightarrow +
\infty$  but does not specify how fast.

For radiation  $(\gamma = \frac{4}{3})$ 
Doroshkevich et al give the following asymptotic
forms
\begin{equation}
\beta^- \approx C_1 T \theta^{- \frac{1}{2}}
\text{sin} [\psi (T)], \quad l^2 e^{-2
\beta^+} \approx C_3 \frac{\theta (T)}{T^2},
\tag{C.13} \label{eqC.13} 
\end{equation}
with
$$
\frac{d \psi}{d T} = C_2 \frac{\theta
(T)}{T^2} , \quad \theta (T) \approx
\frac{1}{ln (\frac{C_4}{T})}, 
$$
where  $C_1, C_2, C_3$  and  $C_4$  are
constants (see Appendix I, equations (I.14 -
I.15); we have relabelled their constants and
have truncated their asymptotic expansions).  In
order to compare with our results we note that 
$$
\frac{d \psi}{d \tau} = \frac{1}{l^3 H} \frac{d
\psi}{dT} \approx \frac{1}{l_0^3 H_0} e^{- \tau}
\frac{d \psi}{dT},
$$
as follows from Table C.1 and equation
\eqref{eqC.10}.  In addition Table C.1 gives  
$T \sim e^{- \tau}$, suppressing constants, and
hence  $\theta \sim \tau^{-1}$  as  $\tau
\rightarrow + \infty$.  It now follows that
\eqref{eqC.13} is in agreement with our result
\eqref{eqC.11}.

Finally, Barrow and Sonoda 1986, using the
equations of Lukash 1975, investigated the
asymptotic behaviour of non-tilted Bianchi
VII$_0$  universes with perfect fluid and
$\gamma$-law equation of state with  $1< \gamma
< \frac{4}{3}$.  They used  $\xi$-time, up to a
factor of 2.  By solving the linearized field
equations they obtained
$$
\beta^- \sim \xi^{-\frac{1}{1-\beta}} (A
\text{cos}  2 \xi + B \text{sin}  2 \xi),
$$
as  $\xi \rightarrow + \infty$, after
compensating for the factor of 2 (see their
equation (4.144)).  This form agrees with our
result \eqref{eqC.10}.  In order to facillitate
comparison, we note that their  $\mu$  is
related to  $\beta^-$  by  $\mu = 2 \sqrt{3}
\beta^-$, and mention that they redefine  $\mu$ 
in their equation (4.138).  Their parameter 
$\alpha$  is related to our  $\beta$  by 
$\alpha = 2 \beta / (1 - \beta)$.  Barrow and
Sonoda argue that their linearization procedure
is consistent only if  $\gamma \leq \frac{5}{4}$,
suggesting that a bifurcation occurs at  $\gamma
= \frac{5}{4}$.  In contrast, we find that a
bifurcation occurs at  $\gamma = \frac{4}{3}$,
and that the asymptotic form \eqref{eqC.10} is
valid for 
$\gamma$  between  $\frac{2}{3}$  and 
$\frac{4}{3}$.

\vspace{4mm}
\section*{Acknowledgements}
We thank Ulf Nilsson for commenting in detail on
the manuscript.  This research was supported by
a grant to J. W. from the Natural
Sciences and Engineering Research Council of
Canada and by an Undergraduate Research Award to
M. H. from the Faculty of Mathematics at
the University of Waterloo.
\vspace {4mm}

\section*{References}

\vspace{3mm}

\noindent
\hangindent=3em
Barrow J D and Sonoda D H 1986 Asymptotic
stability of Bianchi universes {\em Phys. Rep.}
{\bf 139} 1-49

\noindent
\hangindent=3em
Berger B K, Garfinkle D and Strasser E 1997  New
algorithm for Mixmaster dynamics {\em Class.
Quantum Grav.} {\bf 14} L29-L36

\noindent
\hangindent=3em
Carr J 1981 {\em Applications of Center Manifold
Theory} Springer Verlag

\noindent
\hangindent=3em
Collins C B and Hawking S W 1973a The
rotation and distortion of the universe {\em
Mon. Not. R. Astron. Soc.} {\bf 162} 307-20

\noindent
\hangindent=3em
Collins C B and Hawking S W 1973b Why is
the universe isotropic? {\em Astrophys. J.} {\bf
180} 317-34

\noindent
\hangindent=3em
Doroshkevich A G, Lukash V N and Novikov I D
1973 The isotropization of homogeneous
cosmological models {\em Sov. Phys.} JETP
{\bf 37} 739-46

\noindent
\hangindent=3em
Doroshkevich A G, Lukash V N and Novikov I D
1975 Primordial radiation in a homogeneous but
anisotropic universe {\em Sov. Astr.} {\bf 18}
554-60

\noindent
\hangindent=3em
Eardley D 1974 Self-similar
spacetimes: geometry and dynamics {\em Commun.
Math. Phys.} {\bf 37} 287-309

\noindent
\hangindent=3em
Ellis G F R and MacCallum M A H 1969 A
class of homogenous cosmological models,
{\em Commun. Math. Phys.} {\bf 12} 108-41

\noindent
\hangindent=3em
Goode S W 1989  Analysis
of spatially inhomogeneous perturbations of the
FRW cosmologies {\em Phys. Rev. D.} {\bf 39}
2882-92

\noindent
\hangindent=3em
Hirsch M W and Smale S 1974 {\em Differential
Equations, Dynamical Systems, and Linear
Algebra} Academic Press

\noindent
\hangindent=3em
Jantzen R T 1984 Spatially homogeneous dynamics:
a unified picture  In {\em Cosmology of the Early
Universe}, ed. R Ruffini and L Z Fang {\em World
Scientific}

\noindent
\hangindent=3em
LeBlanc V G, Kerr D and Wainwright J 1995 
Asymptotic States of magnetic VI$_0$
cosmologies  {\em Class. Quantum Grav.} {\bf 12}
513-41

\noindent
\hangindent=3em
Lukash V N 1975 Gravitational waves that
conserve the homogeneity of space {\em Sov.
Phys.} JETP {\bf 40} 792-9

\noindent
\hangindent=3em
Maartens R, Ellis G F R and Stoeger W J 1995a
Limits on anisotropy and inhomogeneity from the
cosmic background radiation {\em Phys. Rev. D}
{\bf 51} 1525-35

\noindent
\hangindent=3em
Maartens R, Ellis G F R and Stoeger W J
1995b Improved limits on anisotropy and
inhomogeneity from the cosmic radiation
background radiation {\em Phys. Rev. D} {\bf 51}
5942-5

\noindent
\hangindent=3em
Maartens R, Ellis G F R and Stoeger W J 1996
Anisotropy and inhomogeneity of the
universe from  $\delta T/T$  {\em Astron.
Astrophys.} {\bf 309} L7-L10

\noindent
\hangindent=3em
Misner C W 1969 Quantum cosmology {\em I, Phys.
Rev.} {\bf 186} 1319-27

\noindent
\hangindent=3em
Siklos S T C 1980 Field equations for spatially
homogeneous spacetimes {\em Phys. Lett. A} {\bf
76} 19-21

\noindent
\hangindent=3em
Wainwright J and Ellis G F R (eds) 1997 {\em
Dynamical Systems in Cosmology}  Cambridge
University Press

\noindent
\hangindent=3em
Wainwright J, Coley A A, Ellis G F R and Hancock
M 1998 On the isotropy of the Universe: do
Bianchi VII$_h$  cosmologies isotropize? {\em
Class. Quantum Grav.} {\bf 15} 331-50

\noindent
\hangindent=3em
Weaver M, Isenberg J and Berger B K 1998
Mixmaster Behaviour in Inhomogeneous
Cosmological Spacetimes {\em Phys. Rev. Lett.}
{\bf 80} 2984-7

\noindent
\hangindent=3em
Zel'dovich Ya B and Novikov I D 1983  {\em The
Structure and Evolution of the Universe}
University of Chicago Press

\end{document}